\documentclass[12pt]{article}

\global\arraycolsep=1pt
\usepackage{amsmath}
\usepackage{amssymb}
\newcommand{\appBE}{A}
\newcommand{\be}{\begin{equation}}
\newcommand{\ee}{\end{equation}}
\newcommand{\ba}{\begin{eqnarray}}
\newcommand{\ea}{\end{eqnarray}}
\renewcommand{\theequation}{\thesection.\arabic{equation}}
\newcommand{\Section}{\setcounter{equation}{0} \section}
\newcommand{\Appendix}[2]{
\renewcommand{\theequation}{#1.\arabic{equation}}\setcounter 
{equation}{0}
\renewcommand{\thesubsection}{#1.\arabic{subsection}}\setcounter 
{subsection}{0}
\section*{Appendix #1: #2}
}
\renewcommand{\thefootnote}{\fnsymbol{footnote}}
\newcommand{\bC}{{\mathbb C}}
\newcommand{\bZ}{{\mathbb Z}}
\newcommand{\cN}{{\mathcal N}}

\newcommand{\cW}{{\cal W}}
\newcommand{\ha}{{1\over2}}
\newcommand{\Exp}[1]{\exp\left\{#1\right\}}
\newcommand{\qV}{$q$-Virasoro}

\newcommand{\hw}{h}
\newcommand{\Gaiotto}{G}
\newcommand{\nGaiotto}[1]{G_{#1}}
\newcommand{\tL}{\widehat L}
\newcommand{\px}{p}

\begin{document}

%
\begin{titlepage}
\begin{flushright}
{Oct. 23, 2009 }
\end{flushright}
\vspace{0.5cm}
\begin{center}
{\Large \bf
Five-dimensional AGT Conjecture \\ 
\vskip-3pt
and \\ 
\vskip6pt
the Deformed Virasoro Algebra
}
\vskip10mm
{
{
Hidetoshi Awata}${}^\star$ 
and 
{
Yasuhiko Yamada}${}^\dagger$
}
\vskip 10mm
{\baselineskip=14pt
\it
${}^\star$%
Graduate School of Mathematics \\
Nagoya University, Nagoya 464-8602, Japan\\
\vskip12pt
${}^\dagger$%
Department of Mathematics, Faculty of Science \\
Kobe University, Hyogo 657-8501, Japan
}
\end{center}
\vskip8mm

\begin{abstract}
We study an analog of the AGT (Alday-Gaiotto-Tachikawa) 
relation in five dimensions. 
We conjecture that the instanton partition function of 5D 
$\cN=1$ pure $SU(2)$ gauge theory
coincides with the inner product of the Gaiotto-like state in the  
deformed Virasoro algebra.
In four-dimensional
case, a relation between the Gaiotto construction and the  
theory of Braverman and Etingof is also discussed.
\end{abstract}
\end{titlepage}


\renewcommand{\thefootnote}{\arabic{footnote}} \setcounter 
{footnote}{0}


\Section{Introduction}

In \cite{rf:AGT}, Alday, Gaiotto and Tachikawa discovered
remarkable relations between the 4D $\cN=2$ super  
conformal gauge theories and the 2D Liouville CFT.
This surprising conjecture has been extended 
to asymptotic-free  cases
\cite{rf:Gaiotto,
rf:Ms6} 
and to $SU(n)$-gauge/$\cW_n$-CFT
\cite{rf:Wyllard,
rf:Ms4}. 
Some explanations were given from $M$-theory
\cite{rf:BonelliTanzini,
rf:ABT} 
and from topological string
\cite{rf:DV}. 
Various checks have been done in
\cite{rf:Ms5,
rf:Poghossian} 
for example.
The AGT relation is also closely related with the current active  
subjects such as
loop operators
\cite{rf:DMO}--
\cite{rf:DGOT}, 
wall crossing
\cite{rf:GMN}--
\cite{rf:KS}, 
integrability structures
(see, e.g., \cite{rf:Takasaki,
rf:NS} 
and references therein) and so on.

In this paper, we will examine an extension of the AGT conjecture to  
the 5D gauge theory.
The instanton counting
\cite{rf:Nek}--
\cite{rf:NekrasovOkounkov}
has a natural generalization to the 5D gauge theory 
\cite{rf:NekrasovOkounkov,
rf:NY2} 
that can be viewed as a $q$-analog of 4D cases.
In \cite{rf:AK}--
\cite{rf:AK08} 
it was argued that they are closely related with the
Macdonald symmetric polynomials \cite{rf:Mac}.
On the other hand, 
there exists a natural deformation of the Virasoro/$\cW$ algebra
\cite{rf:SKAO}--%
\cite{rf:AKOS}
which is also related with Macdonald polynomials.
Therefore, 
it is quite natural to expect an extension of the AGT relation
between partition functions of the 5D gauge theory and 
correlation functions of the deformed Virasoro (or $\cW$) algebra. 
In this note, 
we will formulate and check this relation in the simplest case: 
the pure $SU(2)$ gauge theory.
Connections between 5D gauge theories and 2D relativistic
integrable systems have been pointed out in 
\cite{rf:GorskyNekrasov,
rf:Nekrasov96}.

In section 2, following the work by Gaiotto \cite{rf:Gaiotto}, 
we will recall the conjectured relation between
the instanton partition function of 
the 4D $\cN=2$ pure $SU(2)$ gauge theory and 
a Virasoro CFT correlation function
(a two-point function or an inner product) 
with states of ``irregular'' singularity.
In section 3, after a brief review on the deformed Virasoro algebra,  
we will formulate the 5D analog of the AGT conjecture 
in the simplest case: the pure $SU(2)$ gauge theory, and check it. 
We will see a very nontrivial coincidence also in 5D.
Section 4 is devoted to conclusions and discussion.
In the Appendix A, we will make a comment on the relation between 
the AGT conjecture and the instanton counting 
by Braverman and Etingof using the Whittaker-Toda equation
\cite{rf:Braverman,
rf:BravermanEtingof}. 


\section{Review of Gaiotto construction}
\label{secDef}%

In the remarkable work \cite{rf:AGT}, 
Alday, Gaiotto and Tachikawa conjectured the surprising relation 
between the 4D supersymmetric gauge theory 
and the 2D conformal filed theory.
In their original setting, it was conjectured that the instanton  
partition function $Z^{\rm inst}$ of
the 4D $\cN=2$ gauge theories with $N_f=4$ flavours is  
coincides with the conformal
block of the Virasoro CFT with four primary fields. The four fundamental  
matters corresponds
to the four primary fields.
Soon after that, Gaiotto gave extension of this correspondence to the  
theories with $N_f<4$. These theories can be obtained as degeneration 
limit of the $N_f=4$ case and certain confluent type operator/state appear.
The state (which we call Gaiotto state) make irregular singularity
for stress tensor $T(z)$ of order grater than two.
Here, we will recall the construction of Gaiotto in case of $N_f=0$.

Let $V_h$ be the highest weight representation of the Virasoro algebra
\begin{equation}
[L_m, L_n]=(m-n)L_{m+n}+\frac{c}{12}(m^3-m)\delta_{m+n,0},
\label{eq:VirasoroAlgebraMode}%
\end{equation}
generated by the highest weight vector $|h\rangle$ such that
\begin{equation}
L_0 |h\rangle=h |h\rangle, \qquad L_n |h\rangle=0 \quad (n>0).
\end{equation}
The dual representation $V_h^{\ast}$, and the pairing is defined
by;
\begin{equation}
\langle h|L_0=h\langle h|, \qquad \langle h|L_n=0 \quad (n<0),  
\qquad
\langle h|h\rangle=1.
\end{equation}
We will put $c=1+6 (b+1/b)^2$. 


In order to realize the relation $\langle T(z) \rangle=\frac 
{\Lambda^2}{z^3}+\frac{2u}{z^2}+\frac{\Lambda^2}{z}$,
which is the square of the Seiberg-Witten differential,
Gaiotto defined a vector $|\Delta, \Lambda\rangle \in V_\Delta$
by the condition
\be
L_1 |\Delta, \Lambda\rangle=\Lambda^2 |\Delta, \Lambda\rangle,  
\qquad
L_2 |\Delta, \Lambda\rangle=0,
\qquad \Lambda^2\in\bC,
\label{eq:Gaiotto}%
\ee
with leading term normalization 
$|\Delta, \Lambda\rangle=|\Delta\rangle+\cdots$.%
\footnote{
Here $\cdots$ means the higher order terms in $L_0$-degree (level).
}{ }
%
Note that \eqref{eq:VirasoroAlgebraMode} guarantees
$L_n |\Delta, \Lambda\rangle = 0$ for all 
$n$ larger than two.
\footnote{
Note also that if 
$|v\rangle \in V_h$ and $L_n |v\rangle = 0$ for $n=2,3,4$
then $L_n L_1|v\rangle = L_n|v\rangle = 0$
for all $n\geq 2$.
}{ }
Let 
$|\Delta,\Lambda\rangle = \sum_{n=0}^{\infty}\Lambda^{2 n}v_n$
with $L_0 v_n = (\Delta + n) v_n$, $v_0= |\Delta\rangle$,
then \eqref{eq:Gaiotto} is equivalent to
$L_1 v_n = v_{n-1}$ with
$v_{-1} := 0$ and
$L_2 v_n = 0$. 
The uniqueness of $v_n$ is proved by the induction in $n$
using the fact that
the null vector $|\chi\rangle\in V_h$ defined by 
$L_n|\chi\rangle = 0$ for $n\geq 1$
does not exist for the generic $h$ and $c$.

%
The dual vector $\langle \Delta,\Lambda | \in V_\Delta^{\ast}$
is defined similar way:
\begin{equation}
\langle \Delta, \Lambda|L_{-1}=\Lambda^2 \langle\Delta, \Lambda|,
\qquad
\langle \Delta, \Lambda|L_n = 0
\qquad (n\leq -2),
\end{equation}
$\langle \Delta, \Lambda|=\sum_{n=0}^{\infty} \Lambda^{2 n} w_n$,
where $w_n L_0=(\Delta+n) w_n$ and $w_0=\langle \Delta |$.


It is conjectured that the inner product 
$\langle \Delta,\Lambda|\Delta, \Lambda\rangle$
coincides with the instanton part of 
the 4D $\cN=2$ pure $SU(2)$ gauge theory. 
This conjecture has been checked for lower order in 
$\Lambda$-expansion
and proved in the classical limit 
$c \rightarrow \infty$ 
\cite{rf:Ms5}.
In the appendix A, we will give another support on this conjecture by  
using the result of
Braverman and Etingof \cite{rf:BravermanEtingof}.


\section{Five-dimensional case}
\label{secDef}%


\subsection{Quantum deformed Virasoro algebra}


Let $(q,t)$ be formal parameters.
We also let $p:=q/t$ and $t=q^\beta$.
The {\qV} algebra  is an associative algebra generated by
$\langle T_n\rangle_{n\in\bZ}$ with the relation \cite 
{rf:SKAO,rf:AKOSvir},
\be
[T_n \, , \, T_m]=-{\sum^{\infty}_{l=1}}f_l\left(T_{n-l}T_{m+l}-T_ 
{m-l}T_{n+l}\right)
-\frac{(1-q)(1-t^{-1})}{1-p}(p^{n}-p^{-n})\delta_{m+n,0},
\label{eq:qVirAlgebraMode}%
\ee
where $[A,B]=AB-BA$ and
$f(z)={\sum^{\infty}_{l=0}}f_l z^l$ with
\be
f(x)=\exp \Biggl\{ \sum_{n=1}^\infty
{ (1-q^n)(1-t^{-n}) \over 1+p^n} { x^n \over n} \Biggr\}.
\label{eq:structure}%
\ee
The defining relation (\ref{eq:qVirAlgebraMode}) can be written in  
terms of
the current $T(z)=\sum_{n\in\bf{Z}}T_n z^{-n}$ as
\be
f(w/z)T(z)T(w)-T(w)T(z)f(z/w)
=
  -\frac{(1-q)(1-t^{-1})}{1-p} \left\{
        \delta \Bigl(\frac{pw}{z}\Bigr)-
        \delta \Bigl(\frac{w}{pz}\Bigr)\right\},
\label{eq:qVirAlgebraGen}%
\ee
where
$\delta(z)=\sum_{n \in {\bf Z}}z^n$ is the multiplicative 
$\delta$-function
which satisfies $g(z)\delta(z) = g(1)\delta(z)$
for any function $g(z)=\sum_{n\in{\bf Z}} g_n z^n$.




The {\qV} algebra reduces to the Virasoro algebra in the limit 
$q\to 1$.
Let $q=e^{\hbar}$. Then the generator $L(z)$ defined as
\be
T(z) = 2+\beta\hbar^2 
\left\{ z^2 L(z)+{1\over 4}\left(\sqrt{\beta}-{1/ \sqrt{\beta}}\right)^2 \right\}
+ O(\hbar^4),
\label{eq:hbarExpansion}%
\ee
is the ordinary Virasoro current
$ L(z)=\sum_{n \in {\bf Z}}L_{n}z^{-n-2}$, which satisfies
\eqref{eq:VirasoroAlgebraMode}
with the central charge $c=1-6 \left(\sqrt{\beta}-1/\sqrt{\beta}\right)^2$, ($\beta=-b^2$).


\subsection{Deformed Gaiotto state}


%
Let $M_\hw$ be the Verma module over the {\qV} algebra,
generated by the highest weight vector $|\hw\rangle$, such that
$T_n|\hw\rangle=0$ for $n>0$ and
$T_0|\hw\rangle= h |\hw\rangle$ with $\hw\in\bC$.
The dual module $M_\hw^*$ is generated by $\langle\hw|$
which satisfies
$\langle\hw|T_n = 0$ for $n<0$ and
$\langle\hw|T_0= \hw\langle\hw|$.
The bilinear form $M_\hw^*\otimes M_\hw\rightarrow\bC$
is uniquely defined by $\langle\hw|\hw\rangle=1$.
For a generic value of $q$,
the representation theory of the {\qV} algebra is similar to
that of the Virasoro algebra \cite{rf:BP}.
%
%
Let us introduce the (outer) grading  operator $d$ which satisfies
$[d , T_n]= n T_n$ and set $d|\hw\rangle=0$. We call a vector
$|v\rangle \in M(\hw)$ of level $n$ if $d|v\rangle=-n|v\rangle$.
Then $M_\hw$ has the grade decomposition
$M_\hw = \oplus_{n=0}^\infty M_{\hw,n}$ by the grade $n$ subspace  
$M_{\hw,n}$.


The deformed analog of the Gaiotto state 
$|\Gaiotto\rangle\in M_\hw$ is defined by 
\be
T_1|\Gaiotto\rangle = \Lambda^2|\Gaiotto\rangle,
\qquad
T_n|\Gaiotto\rangle = 0\quad (n\geq 2),
\qquad \Lambda^2\in\bC,
\label{eq:qGaiotto}%
\ee
with a normalization condition
$|G \rangle = |h\rangle + \cdots$.
Let $|\Gaiotto\rangle = 
\sum_{n=0}^\infty \Lambda^{2n}|\nGaiotto  n\rangle$
with $|\nGaiotto n\rangle\in M_{\hw,n}$,
then \eqref{eq:qGaiotto} is equivalent to
$T_1|\nGaiotto n\rangle = |\nGaiotto {n-1}\rangle$ with
$|\nGaiotto {-1}\rangle := 0$ and
$T_m|\nGaiotto n\rangle = 0$ for $m\geq 2$.
The dual vector $\langle G | \in M_h^{\ast}$ 
is defined similar way:
\be
\langle G|T_{-1}=\Lambda^2 \langle G|,
\qquad
\langle G|T_n = 0
\qquad (n\leq -2),
\label{eq:dualqGaiotto}%
\ee
and $\langle G| = \langle h| + \cdots$.
To calculate an inner product
$\langle\Gaiotto\vert\Gaiotto\rangle
=
\sum_{n=0}^\infty \Lambda^{4n}\langle\nGaiotto n|\nGaiotto n 
\rangle$,
it is convenient to introduce a free boson realization for
the {\qV} algebra.


\subsection{Free boson realization}


The {\qV} algebra has a free field representation:
\begin{eqnarray}
T(z)&=&\sum_{n \in {\bf Z}} T_n z^{-n}=\Lambda_1(z)+\Lambda_2(z), 
\\
\Lambda_1(z)&=&
\Exp{-\sum_{n=1}^{\infty} t^{n}a_{-n}z^n}
\Exp{\sum_{n=1}^{\infty} q^n a_{n} z^{-n}}K, \\
\Lambda_2(z)&=&
\Exp{\sum_{n=1}^{\infty} q^{n}a_{-n}z^n}
\Exp{-\sum_{n=1}^{\infty} t^n a_{n} z^{-n}}K^{-1}.
\end{eqnarray}
Where $a_n$ are the Heisenberg operators realized as
\begin{equation}
a_n=\frac{q^n-1}{t^n} \frac{\partial}{\partial \px_n}, 
\qquad 
a_{-n}= v_n \px_n,
\qquad
v_n:=\frac{t^{n}-1}{n(t^n+q^n)q^n}
\qquad (n>0),
\label{eq:qVirBoson}%
\end{equation}
with the variables $\px=(\px_1, \px_2,\cdots)$
and $K$ is a multiplicative charge operator which commute with $a_n$'s.
Let $|k\rangle$ be the highest weight vector such as
$K|k\rangle=k|k\rangle$ and
$a_n |k\rangle=0$ $(n>0)$.
Then
$T_0 |k\rangle = (k+k^{-1})|k\rangle$.
The deformed Gaiotto vector is defined by \eqref{eq:qGaiotto} and 
$|G\rangle=|k\rangle+\cdots$.
This over determined system of linear equations
have unique solution.%
\footnote{
If $v_n$'s are considered as free parameters 
without the relations \eqref{eq:qVirBoson}, 
the equations become inconsistent at level 5. 
It may be interesting to study the general form of
the parameters $v_n$ which admits the Gaiotto like state.
}{ }
The vector $|G\rangle$ is realized as a polynomial in
$\px=(\px_1, \px_2,\cdots)$%
\begin{equation}
G(\px)=1+\Lambda^2 c_1 \px_1
+\Lambda^4 (c_{11}\px_1^2+c_{2}\px_2)+\cdots,
\end{equation}
The first few coefficients are as follows;
\begin{eqnarray}
c_1
&=&
\frac{kt}{(k^2 q-t)(q-1)}
,\nonumber\\
c_{11}
&=&
\frac{k^2 (k q-t) (k q+t)t^2}
{2 
\left(k^2 q^2-t\right) 
\left(k^2 q-t\right) 
\left(k^2 q-t^2\right)
(q-1)^2}
,\nonumber\\
c_2
&=&
-
\frac{k^2 \left(k^2 q^2+t^2\right)t^2}
{2 
\left(k^2 q^2-t\right) 
\left(k^2 q-t\right)
\left(k^2 q-t^2\right)
(q^2-1)}
,\nonumber\\
c_{111}
&=&
\frac{k^3 \left(k^4 q^5-2 k^2 t^3 q^3+k^2 t^2 q^3+k^2 t^3
q^2-2 k^2 t^2 q^2+t^5\right)t^3}
{6 
\left(k^2 q^3-t\right) 
\left(k^2 q^2-t\right) 
\left(k^2 q-t\right) 
\left(k^2 q-t^2\right)  
\left(k^2 q-t^3\right)
(q-1)^3}
,\nonumber\\
c_{21}
&=&
-
\frac
{k^3 \left(k^2 q^3-t^3 \right)\left(k^2 q^2+t^2\right)t^3}
{2
\left(k^2 q^3-t\right) 
\left(k^2 q^2-t\right) 
\left(k^2 q-t\right) 
\left(k^2 q-t^2\right)  
\left(k^2 q-t^3\right)
(q-1)(q^2-1)}
,\nonumber\\
c_{3}
&=&
\frac{k^3 \left(k^4 q^5+k^2 t^3 q^3+k^2 t^2 q^3+k^2 t^3 q^2 
+k^2 t^2 q^2+t^5\right)t^3}
{3 
\left(k^2 q^3-t\right) 
\left(k^2 q^2-t\right) 
\left(k^2 q-t\right) 
\left(k^2 q-t^2\right)  
\left(k^2 q-t^3\right)
(q^3-1)}
.
\end{eqnarray}

As in \cite{rf:SKAO}, one can identify the boson Fock space with the
space of symmetric functions in variables, say $x=(x_1,x_2,\cdots)$,
by the natural identification $p_n=\sum_i x_i^n$. 
Then the function $G(p)$ is written as a sum of the Macdonald function 
$P_{\lambda}(x;q,t)$ \cite{rf:Mac} as 
\be
G(\px) =
\sum_\lambda \Lambda^{2|\lambda|} P_\lambda(x;q,t)
\prod_{(i,j)\in\lambda}
{k\over 1-k^2 q^j  t^{-i}}
{q^{\lambda_i-j}\over 1-q^{\lambda_i-j+1}t^{\lambda^\vee_j-i}},
\ee
which we have checked for $|\lambda|\leq 8$.
Similar structure can be observed in 4D case in terms of the Jack polynomials.

Next, we compute the dual vector $\langle G|$ defined by \eqref{eq:dualqGaiotto}.
To do this, we use an involution $\iota$ such that
\begin{equation}
\iota(AB)=\iota(B)\iota(A), \quad \iota(T_n)=T_{-n}, \quad
\iota(a_n)= 
a_{-n}, \quad \iota(K)=K^{-1}.
\end{equation}
Then, the free field realization of the vector 
$\langle G|$ is obtained from the polynomial $G(\px)$
simply by the replacement
\begin{equation}
\px_n \rightarrow 
{u_n}\frac{\partial}{\partial \px_n}, 
\qquad
u_n:=-n(t^n+q^n)
\left(\frac qt \right)^n
\frac{1-q^n}{1-t^{n}} 
,
\qquad
k \rightarrow k^{-1}.
\end{equation}
Hence the inner product is given as
\begin{equation}
\langle G | G\rangle =\sum_{\lambda}\Lambda^{4 |\lambda|}
c_{\lambda}(k)c_{\lambda}(k^{-1})w_{\lambda},
\end{equation}
where for a partition
$\lambda=(\lambda_1,\lambda_2,\cdots,\lambda_\ell)=
(1^{m_1}2^{m_2}\cdots |\lambda|^{m_{|\lambda|}})$,
\begin{equation}
w_{\lambda}:=
\prod_{i=1}^{|\lambda|}
{m_i!}{u_i^{m_i}} .
\end{equation}
Then we have
\begin{equation}
\langle G | G\rangle=1+\Lambda^4 k^2 \frac{qt(q+t)}{
(q-1) 
\left(k^2 q-t\right) 
\left(q-t k^2\right)
(1-t)
}+\Lambda^8 k^4 (qt)^2 \frac{A}{B}+\cdots,
\end{equation}
where
\begin{equation}
\begin{array}{rl}
A=
&
k^2 q^7
+k^2 t^2 q^6+2 k^2 t q^6
-k^4 t^3 q^5-t^3 q^5+2 k^2 t^2q^5+k^2 t q^5
\\
&
-k^4 t^4 q^4-t^4 q^4-k^4 t^3 q^4+k^2 t^3 q^4-t^3 q^4
-k^4 t^2 q^4-t^2 q^4
\\
&
-k^4 t^5q^3-t^5 q^3-k^4 t^4q^3+k^2 t^4 q^3-t^4 q^3
-k^4 t^3 q^3-t^3 q^3
\\
&
+k^2 t^6 q^2+2 k^2 t^5q^2-k^4 t^4 q^2-t^4 q^2
+2 k^2 t^6 q+k^2 t^5 q
+k^2 t^7, \\
B=&
(q^2-1)(q-1) 
\left(k^2 q^2-t\right) 
\left(k^2 q-t\right) 
\left(k^2 q-t^2\right) 
(1-t)
(1-t^2) 
\\
&\hskip66pt\times
\left(q^2-t k^2\right) 
\left(q-t k^2\right) 
\left(q-t^2 k^2\right)
.
\end{array}
\end{equation}
This is the 5D analog of $\langle \Delta, \Lambda | \Delta,  
\Lambda \rangle$
which can be compared with the 5D instanton partition function.




The instanton part of the Nekrasov's five-dimensional 
pure $SU(2)$ partition function
is written by a sum over two Young diagrams $\lambda$ and $\mu$ 
as follows \cite{rf:NY2};%
\footnote{
The parameter $Q$ is related 
with the vacuum expectation value $a$ of the scalar fields in the  
vector multiplets as $Q=q^a$
and the parameters $(q,t)$ are related with those
$(\epsilon_1,\epsilon_2)$ of the $\Omega$ background  
through $(q,t)=(e^{\epsilon_2},e^{-\epsilon_1})$.
}{ }
\ba
Z^{\rm inst}
&=&
\sum_{\lambda, \mu}
{
\left(\Lambda^4 t/q \right)^{|\lambda|+|\mu|}
\over
N_{\lambda\lambda}(1)
N_{\lambda\mu}(Q)
N_{\mu\mu}(1)
N_{\mu\lambda}(Q^{-1})
},
\cr
N_{\lambda\mu}(Q)
&:=&
\prod_{(i,j)\in\mu }
\left( 1 - Q\, q^{\lambda_i-j} t^{\mu^\vee_j-i+1} \right)
\prod_{(i,j)\in\lambda}
\left( 1 - Q\, q^{-\mu_i+j-1} t^{-\lambda^\vee_j+i  } \right).
\label{eq:Nek}%
\ea
Here, $\lambda = (\lambda_1,\lambda_2,\cdots)$ is a Young diagram/partition
such that $\lambda_{i} \geq \lambda_{i+1}$. 
%
$\lambda^\vee $ is its conjugate (dual) as Young diagram
and $|\lambda| = \sum_i \lambda_i$.
Then we observe the following coincidence:
\be
Z^{\rm inst} =  \langle\Gaiotto\vert\Gaiotto\rangle,
\ee
for $k=Q^\ha$. We have checked this up to level 9. 


\section{Summary and discussion}


We studied 5D analog of AGT conjecture in 
the simplest pure $SU(2)$ gauge theory.
It turns out that a natural generalization of the Gaiotto  
construction using the deformed Virasoro algebra
is related to the gauge theory. 
Fortunately, the parameters $(q,t)$ are directly related 
in both sides of the correspondence. 
This is because these parameters have already been chosen 
in order to fit the conventions in Macdonald polynomials. 
It is quite natural to expect similar relation between $SU(n)$ gauge
theory and the deformed $\cW_n$ algebras
\cite{rf:FFr,
rf:AKOS}.
Inclusion of matter fields is also interesting problem.
For the proof of the conjecture, the vertex operator constructions
\cite{rf:NekrasovOkounkov,
rf:MNTT,
rf:Carlsson} would be useful.

In 4D case, the Gaiotto states produce irregular singularities.
Such a conformal field theory with irregular singularities 
\cite{rf:JNS} are recently studied from the point of view
of quantum Painlev\'e equations. 
{}From this point of view, the AGT relation may
be considered as a quantization of 
the correspondence between the Seiberg-Witten theory
and the Hitchin/Painlev\'e equations 
(see \cite{rf:KMNOY} for example).


\section*{Acknowledgments}


We would like thank  H. Kanno, H. Nakajima and Y. Tachikawa for  
discussions.
The work of Y.Y. is supported in part by Grant-in-Aid for Scientific  
Research
[\#21340036] from the Japan Ministry of Education, Culture, Sports,  
Science and Technology.
The work of H.A. is supported in part by Daiko Foundation.

\Appendix{\appBE}{Relation to the Braverman and Etingof}


We follow the notation in section 2. 
Let us consider the one-point function
\begin{equation}
\Psi(z)=
\langle\Delta_2, \Lambda|\Phi_h(z)|\Delta_1, \Lambda\rangle.
\end{equation}
We will derive the 2nd order differential equation for $\Psi(z)$ 
when the operator $\Phi_h(z)$ is the degenerate primary field 
of conformal dimension $h=h_{2,1}$.
To do this, we consider the insertion of the stress tensor $T(w)$:
\begin{equation}
\langle \Delta_2, \Lambda|T(w)\Phi_h(z)|\Delta_1, \Lambda\rangle
=\sum_{n \in {\bf Z}} \langle \Delta_2, \Lambda|L_n
w^{-n-2}\Phi_h(z)|\Delta_1, \Lambda\rangle.
\end{equation}
Using the defining relations of the states $\langle \Delta_2,
\Lambda|$, $|\Delta_1, \Lambda\rangle$ 
and the operator $\Phi_h(z)$ :
\begin{equation}
[L_n, \Phi_h(z)]=z^n\Big[z\frac{d}{dz}+h(n+1)\Big]\Phi_h(z),
\end{equation}
one can derive the following relation
\ba
\langle \Delta_2, \Lambda|T(w)\Phi_h(z)|\Delta_1, \Lambda\rangle
&=&
\Big[\frac{\Lambda^2}{w}+\frac{\Lambda^2}{w^3}+(\frac{1}{w-z}- 
\frac{1}{w})\frac{\partial}{\partial z}
+\frac{h}{(w-z)^2}\Big]\Psi(z)
\cr
&&\hskip24pt
+\frac{1}{w^2} \langle \Delta_2, \Lambda|\Phi_h(z)L_0|\Delta_1,
\Lambda\rangle.
\ea
On the other hand, by noting the relation
\begin{equation}
\frac{1}{2}\Lambda \frac{\partial}{\partial \Lambda}|\Delta_1,
\Lambda\rangle=(L_0-\Delta_1)|\Delta_1, \Lambda\rangle, \qquad
\frac{1}{2}\Lambda \frac{\partial}{\partial \Lambda}\langle 
\Delta_2,
\Lambda|=\langle\Delta_2, \Lambda|(L_0-\Delta_2),
\end{equation}
we have
\begin{equation}
\langle \Delta_2, \Lambda|\Phi_h(z)L_0|\Delta_1,
\Lambda\rangle=\Big[\frac{\Lambda}{4}\frac{\partial}{\partial 
\Lambda}+
\frac{\Delta_1+\Delta_2-h}{2}-\frac{z}{2}\frac{\partial} 
{\partial z}\Big]\Psi(z).
\end{equation}
Thus we get the following formula
\ba
\langle\Delta_2,\Lambda|T(w)\Phi_h(z)|\Delta_1,\Lambda\rangle
&=&
\Big[\frac{\Lambda^2}{w}+\frac{\Lambda^2}{w^3}
+(\frac{1}{w-z}- \frac{1}{w})\frac{\partial}{\partial z}
+\frac{h}{(w-z)^2}
\cr 
&& 
+\frac{1}{w^2}\Big(
\frac{\Lambda}{4}\frac{\partial}{\partial \Lambda}
+\frac{\Delta_1+\Delta_2-h}{2}
-\frac{z}{2}\frac{\partial}{\partial z}\Big)
\Big]\Psi(z).
\ea
This gives the correlation function 
of the 2nd descendant operator
${\tL_{-2}}\Phi_h(z)$
defined by 
\be
T(w)\Phi_h(z)
=\sum_{n \in {\bf Z}} 
(w-z)^{-n-2} \tL_n \Phi_h(z),
\ee
as follows
\ba
&& \hskip-12pt
\langle\Delta_2,\Lambda|{\tL_{-2}}\Phi_h(z)
|\Delta_1,\Lambda\rangle
=
\frac{1}{z^2}\Big[{\Lambda^2}(z+\frac{1}{z})-\frac{3}{2}z\frac 
{\partial}{\partial z}
+\frac{\Lambda}{4}\frac{\partial}{\partial \Lambda}+
\frac{\Delta_1+\Delta_2-h}{2} \Big]\Psi(z).
\ea
When the primary field $\Phi_h$ has the dimension
\begin{equation}
h=h_{2,1}=-\frac{1}{2}-\frac{3}{4 b^2},
\end{equation}
and the null vector at level 2
\begin{equation}
(b^2 \tL_{-1}^2+\tL_{-2})\Phi_h(z)=0,
\end{equation}
we obtain the following differential equation
\begin{equation}
\Big[b^2 z^2 \frac{\partial^2}{\partial z^2}
+{\Lambda^2}(z+\frac{1}{z})-\frac{3}{2}z\frac{\partial}{\partial  
z}
+\frac{\Lambda}{4}\frac{\partial}{\partial \Lambda}+
\frac{\Delta_1+\Delta_2-h}{2}\Big]\Psi(z)=0.
\end{equation}
According to the fusion rule 
$[h_{2,1}]\otimes[\Delta(a)]=[\Delta(a\pm \frac{1}{2b})]$
where $\Delta(a)=(b+1/b)^2/4-a^2 $, 
we will take
$\Delta_1=\Delta(a+\frac{1}{4b})$ and $\Delta_2=\Delta(a-\frac{1}{4b})$.
Then, by putting $\Psi(z)=z^{\Delta_2-\Delta_1-h_{2,1}}Y(x)$ and $z=e^x$, 
we finally obtain the following differential equation for $Y(x)$
\begin{equation}
\Big[b^2 \frac{\partial^2}{\partial x^2}
+2 ab \frac{\partial}{\partial x}
+{\Lambda^2}(e^x+e^{-x})
+\frac{\Lambda}{4}\frac{\partial}{\partial\Lambda}
\Big]Y(x)=0.
\end{equation}
This equation coincides with the differential equation of
Braverman-Etingof (eq(2.10) \cite{rf:BravermanEtingof})
whose WKB like solution%
\footnote{
The WKB parameter $\kappa$ in front of
$\Lambda \frac{\partial}{\partial \Lambda}$ is hidden in our normalization.
}{ }
gives the instanton part of the 4D $\cN=2$ pure $SU(2)$ gauge theory
with equivalent parameters
$(\epsilon_1, \epsilon_2)=(b,0)$.
Note that in the AGT conjecture, one put 
$(\epsilon_1,\epsilon_2)=(b,\frac{1}{b})$. 
The difference corresponds to the charge shift in 
\hbox{$\langle\Delta(a-\frac{1}{4b}),\Lambda|
\Phi_h(z)|\Delta(a+\frac{1}{4b}),\Lambda\rangle$}.
It will be interesting to study the 5D difference analog 
of these differential equations,
which may be related with the relativistic Toda system
\cite{rf:GorskyNekrasov,
rf:Nekrasov96}.




\end{document}